\documentclass[a4paper,11pt,preprintnumbers]{article}
\pdfoutput=1 

\usepackage{jcappub} 
\usepackage{mathabx}
\usepackage{mathtools}
\usepackage{appendix}
\usepackage{slashed}
\usepackage{multirow}
\usepackage{array}
\usepackage{cellspace}
\usepackage{xcolor}
\usepackage{makecell}
\usepackage{braket}
\usepackage{amsmath, amsfonts, amssymb}
\usepackage{mathrsfs}
\usepackage{subcaption}
\usepackage[normalem]{ulem}

\usepackage{aasmacros}

\usepackage[T1]{fontenc} 

\newcommand{\MPBH}{{M}}
\newcommand{\MPBHini}{{M}_{\rm in}}
\newcommand{\Tform}{T_{\rm form}}

\newcommand{\TBH}{T_{\rm BH}}

\newcommand{\rhotot}{\rho_{\rm tot}}
\newcommand{\rhorad}{\rho_{\rm rad}}

\newcommand{\mycomment}[1]{}

\newcommand{\rcrit}{r_{\rm core}}
\newcommand{\rdec}{r_{\rm dec}}

\newcommand{\Tcore}{T_{\rm core}}

\newcommand{\rcore}{r_{\rm core}}

\title{Neutrino Emission and Plasma Heating from Primordial Black Holes: An Improved Approach to $N_{\mathrm{eff}}$ Constraints}

\author[1]{Héctor Sanchis}
\author[1]{, Gabriela Barenboim}
\author[2]{, Yuber F. Perez-Gonzalez}

\affiliation[1]{Departament de Física Teórica and IFIC, Universitat de València-CSIC, E-46100, Burjassot, Spain}
\affiliation[2]{Departamento de Física Teórica and Instituto de Física Teórica UAM/CSIC,\\
Universidad Autónoma de Madrid, Cantoblanco, 28049 Madrid, Spain}

\emailAdd{hector.sanchis@uv.es}
\emailAdd{gabriela.barenboim@uv.es}
\emailAdd{yuber.perez@uam.es}

\abstract{
    We investigate the impact of neutrino emission via Hawking radiation from primordial black holes (PBHs) on the cosmological effective number of neutrino species, $N_{\mathrm{eff}}$, after neutrino decoupling. By comparing this effect with observational limits, we derive bounds on the abundance of light PBHs. Our analysis incorporates two previously unaccounted-for effects: the emission of secondary neutrinos from unstable particles, which increases $N_{\mathrm{eff}}$, and the modification of the neutrino-photon temperature ratio due to particle emission heating the photon plasma, which lowers $N_{\mathrm{eff}}$. Overall, including these effects allows us to impose constraints on PBHs with initial masses in the range $10^9~{\rm g}\lesssim M_{\rm ini} \lesssim 10^{13}~{\rm g}$. However, our limits remain less stringent than those derived from Big Bang Nucleosynthesis.
}

\preprintnumber{\newline IFT-UAM/CSIC-25-19}

\begin{document}

\maketitle
\flushbottom

\section{Introduction}

Black holes have been a subject of scientific interest for over a century, yet many fundamental questions about them remain unresolved. While the formation of stellar-mass black holes from the collapse of massive stars is well understood, the origin of the supermassive black holes observed at the centers of galaxies remains an open question~\cite{SMBH_Origins_2021NatRP...3..732V}.  

It is possible that black holes formed in the early universe, giving rise to what are known as primordial black holes (PBHs). 
A key property of black holes is that they are expected to emit thermal radiation~\cite{Hawking:1974rv, Hawking:1975vcx}. 
This Hawking radiation includes all degrees of freedom existing in nature. 
However, only low-mass black holes emit significant amounts of radiation, making the detection of Hawking radiation from known astrophysical black holes unlikely. Light PBHs, if they exist, could emit detectable levels of Hawking radiation.  
Depending on their initial mass, strong constraints exist on the abundance of PBHs that could have existed or still persist in the Universe~\cite{Carr:2020gox,Carr:2020xqk,Auffinger:2022khh}.
These objects have been proposed as a potential explanation for various astrophysical and cosmological mysteries, including the origin of supermassive black holes~\cite{SMBH_Origins_2021NatRP...3..732V}, the nature of dark matter~\cite{Khlopov:2008qy,Lennon:2017tqq,Morrison:2018xla,Hooper:2019gtx,Auffinger:2020afu,Gondolo:2020uqv,Bernal:2020bjf,Bernal:2020ili,Bernal:2020kse,Baldes:2020nuv,Masina:2020xhk,Masina:2021zpu,Sandick:2021gew,Bernal:2021bbv,Bernal:2021yyb,Cheek:2021odj,Cheek:2021cfe,Barman:2021ost,Bernal:2022oha,Cheek:2022mmy,Chen:2023lnj,Chen:2023tzd,Kim:2023ixo,Haque:2023awl,RiajulHaque:2023cqe,Chaudhuri:2023aiv,Gehrman:2023esa,Gehrman:2023qjn,Carr:2016drx,Clesse:2016vqa,Carr:2017jsz,Green:2020jor,Croon:2020ouk} and the matter-antimatter asymmetry in the universe~\cite{Barrow:1990he,Majumdar:1995yr,Upadhyay:1999vk,Dolgov:2000ht,Bugaev:2001xr,Baumann:2007yr,Hooper:2020otu,Gehrman:2022imk,Fujita:2014hha,Perez-Gonzalez:2020vnz,Bernal:2022pue,JyotiDas:2021shi,Calabrese:2023key,Calabrese:2023bxz,Schmitz:2023pfy,Ghoshal:2023fno,Datta:2020bht,Gunn:2024xaq,Calabrese:2025sfh, Hamada:2016jnq}. Moreover, PBHs could also generate observable gravitational waves~\cite{Papanikolaou:2020qtd,Domenech:2020ssp,Bhaumik:2022pil,Bhaumik:2022zdd,Ghoshal:2023sfa} or other forms of dark radiation~\cite{Hooper:2019gtx,Lunardini:2019zob,Masina:2020xhk,Masina:2021zpu,Hooper:2020evu,Arbey:2021ysg,Cheek:2022dbx, Wu:2024uxa, Das:2023oph} and affect the stability of the Higgs potential~\cite{Burda:2016mou,Burda:2015isa,Hamaide:2023ayu}.

This work explores the impact of Hawking radiation from light PBHs—particularly neutrino emission—on the effective number of relativistic species, $N_{\rm eff}$, and its gravitational effects on the universe. Since these effects are observationally constrained, they provide a means to place bounds on the abundance of light PBHs.  
Our approach primarily focuses on modifications to the total energy density of neutrinos and photons. However, the injection of high-energy neutrinos could also induce spectral distortions in the Cosmic Microwave Background (CMB). Accurately determining such effects would require solving the Boltzmann equations. While recent studies have introduced novel approaches to address these challenges, see e.g.~\cite{Ovchynnikov:2024xyd,Ovchynnikov:2024rfu,Akita:2024nam,Akita:2024ork}, in this work, we restrict our analysis to the impact of PBH evaporation on the total energy density.

This approach has been previously studied in Refs.~\cite{Hooper:2019gtx,Lunardini:2019zob,Masina:2020xhk,Masina:2021zpu}, but here we incorporate additional physical effects that are relevant for neutrino emission after neutrino decoupling. 
Specifically, we account for: (i) secondary neutrinos produced from the decay of other Hawking-emitted particles, which enhance $N_{\mathrm{eff}}$ and strengthen the bounds, and (ii) the heating of the photon plasma by non-neutrino Hawking radiation, which modifies the neutrino-photon temperature ratio and weakens the bounds. These effects do not merely shift existing constraints; they also provide deeper insights into the underlying physical processes.  

Our results demonstrate that incorporating these effects leads to tighter bounds on PBH abundance through this mechanism, though these constraints remain weaker than those derived from other cosmological probes, such as Big Bang Nucleosynthesis (BBN)~\cite{Kohri:1999ex,Keith:2020jww,Acharya:2020jbv,Boccia:2024nly}. Furthermore, we find that heating the cosmic plasma can effectively reduce the measurable gravitational impact of neutrinos, allowing for a greater number of extra neutrinos without violating observational constraints. 
This insight could have implications beyond the scope of this study.  

The paper is structured as follows: Sec.~\ref{sec:PBHReview} provides a brief overview of PBHs, their Hawking radiation, and existing observational constraints. Sec.~\ref{sec:PBHBoundsNeff} details our method for deriving new bounds on PBH abundance and compares it to previous approaches. The resulting bounds are presented and analyzed in Sec.~\ref{sec:Results}. Finally, we summarize our conclusions in Sec.~\ref{sec:Conclusions}.
We include three appendices: App.~\ref{ap:EnergyDependentNeutrinoDecoupling}, where we describe a procedure to determine whether neutrinos from evaporation would interact with the plasma, App.~\ref{ap:MuonsPionsKaons}, where we describe the calculation needed to determine whether weakly-decaying emitted particles such as muons, pions and kaons decay before getting to thermalize or thermalize before decaying, and App.~\ref{ap:SpectraCalibration}, which considers the calibration that we used for the Hawking spectra from {\tt BlackHawk}~\cite{Arbey_2019_BlackHawk,Arbey:2021ysg}.
Throughout this work, we adopt natural units ($\hbar = c = k_B = 1$), unless stated otherwise.

\section{A short review on PBHs} \label{sec:PBHReview}

To form black holes with masses significantly smaller than the Chandrasekhar limit, typically associated with astrophysical black holes, it is necessary for overdensities to exceed a critical threshold since in a radiation dominated Universe, the pressure would prevent the gravitational collapse~\cite{Carr:2020gox,Carr:2020xqk}.
In general, there are several theoretical scenarios have been proposed to explain how such large density perturbations could develop, leading to the formation of primordial black holes (PBHs).
In this work, we assume that a population of PBHs formed due to some unspecified mechanism, and it is characterized by its initial mass, $\MPBHini$, i.e., we consider a monochromatic mass distribution, and initial energy density, $\beta^\prime$. 
In an initially radiation-dominated Universe, the initial PBH mass can be related to the plasma temperature as follows,
\begin{align}\label{eq:Min}
 \MPBHini &= \frac{4\pi}{3}\, \gamma\, \frac{\rhorad(\Tform)}{H^3(\Tform)}\sim 2\times10^{12}~{\rm g} \left(\frac{\gamma}{0.2}\right) \left(\frac{10^{11}~{\rm GeV}}{\Tform}\right)^2\,.
\end{align}
Here $\rhorad$ represents the radiation energy density at the black hole formation temperature $\Tform$, while $H$ denotes the Hubble rate, and $\gamma \sim 0.2$ defines the gravitational collapse factor for a radiation-dominated Universe.
Inspecting the Schwarzschild radius for these PBH,
\begin{align}
 r_{\rm S, in} &= 2 G\MPBHini\sim 2.7 \times 10^{-5}~{\rm fm} \left(\frac{\MPBHini}{10^{10}~{\rm g}}\right)\,,
\end{align}
we observe that these black holes are microscopic, so that we need to consider how quantum effects affect the evolution of these objects.

Building on this fact, and applying a semi-classical approximation in which quantum fields propagate within a classical gravitational field, Hawking~\cite{Hawking:1974rv,Hawking:1975vcx} demonstrated that black holes emit radiation across all existing degrees of freedom in nature. The differential emission rate per unit energy $E$, per unit time $t$, for a particle of type $i$ with mass $m_i$ is given by
\begin{equation}\label{eq:HawkSpec}
 \frac{d^{2}N_i}{dt\, dE} = \frac{g_i}{2\pi}\frac{\vartheta(\MPBH,E)}{e^{E/T_{{\rm BH}}}-(-1)^{2s_i}}\,
\end{equation}
with $g_i$ are the internal degrees of freedom of the particle, and, more importantly, the temperature associated to the black hole is
\begin{align}\label{eq:HawkTemp}
 \TBH = \frac{1}{8\pi\, G \MPBH}\,\sim 1~{\rm TeV} \left(\frac{10^{10}~g}{\MPBH}\right),
\end{align}
where $M$ is the instantaneous PBH mass.
In particular, the dependence of the emission rate on the internal degrees of freedom of the particle implies that the amount of neutrinos emitted as Hawking radiation will depend on whether neutrinos are Dirac or Majorana particles. More specifically, in the Dirac case, the amount of neutrinos that are directly emitted from a black hole (so-called primary neutrinos) will be twice as much as in the Majorana case, due to the presence of right-handed neutrinos. In contrast, neutrinos that come from the decays of other emitted particles (so-called secondary neutrinos) always come from weak decays, which only produce left-handed neutrinos in any case, so their abundance is the same in both cases~\cite{Lunardini:2019zob}.

The Hawking spectrum in Eq.~\eqref{eq:HawkSpec} departs from being the spectrum of a blackbody due to the presence of the spin-dependent absorption probability $\vartheta(\MPBH,E)$. 
Since the curved spacetime around a black hole effectively acts as a potential barrier, a particle coming from infinity could be either back-scattered or absorbed by the black hole.
Such an absorption probability, given by $\vartheta(\MPBH,E)$, is the same as the probability of an emitted particle to reach spatial infinity.

As real particles are emitted from the black hole's gravitational field, it gradually loses mass. The rate of this mass loss can be estimated by~\cite{Hawking:1975vcx, MacGibbon:1990zk, MacGibbon:1991tj}
\begin{align} \label{eq:M_lossrate}
 \frac{dM}{dt} = -\sum_i \int_{M_i}^{\infty} \frac{d^{2}N_i}{dt\, dE}\, E\, dE = - \varepsilon(M)\, \frac{1}{G^2 M^2}\,,
\end{align}
where $G$ is the gravitational constant.
Here, we have introduced the mass evaporation function $\varepsilon(M)$, which captures the degrees of freedom that can be emitted at any given moment, corresponding to the instantaneous mass of the black hole. We adopt the parametrization of this function as outlined in Ref.~\cite{Cheek:2021odj}.
Crucially, the estimation above assumes that the gravitational field remains largely unaffected by the emission of a particle. However, this assumption eventually breaks down, requiring consideration of the backreaction to the metric caused by the Hawking emission process. Some studies have incorporated this backreaction using the semi-classical approximation, showing that the mass loss rate still follows the $M^{-2}$ law as given in Eq~\eqref{eq:M_lossrate}~\cite{Bardeen:1981zz,Massar:1994iy,Brout:1995rd}.

Thus, naively, one would expect the Hawking spectrum and the mass loss rate equations to remain valid until the black hole's mass reaches the Planck mass~\cite{Hawking:1975vcx}, where quantum gravity effects should be relevant.
Nevertheless, when examining the time evolution of the von Neumann entropy of Hawking radiation alongside the coarse-grained black hole entropy, assumed to match the Bekenstein-Hawking value~\cite{Bekenstein:1972tm,Bekenstein:1973ur}, Page observed that the von Neumann entropy exceeds the black hole's available degrees of freedom at a point known as the Page time, leading to the information paradox~\cite{Hawking:1976ra,Almheiri:2020cfm,Buoninfante:2021ijy, Perez-Gonzalez:2025try}. 
Despite recent progress in addressing the paradox~\cite{Almheiri:2020cfm}, several unresolved questions remain regarding the thermal characteristics of Hawking evaporation beyond the Page time and the potential modifications to the mass loss rate.
In light of the uncertainty surrounding these possible alterations, we will assume that the semi-classical approximation holds true up to approximately the Planck scale, enabling the primordial black hole time evolution to conform to the mass loss rate specified in Eq.~\eqref{eq:M_lossrate}.

The second parameter that characterizes the PBH population is the initial energy density $\rho_{\rm BH}(\Tform)$.
Following Ref.~\cite{Carr:2020gox}, we define such initial PBH abundance through the parameter $\beta^\prime$, that relates the total and PBH energy densities via
\begin{align}\label{eq:beta}
 \beta^\prime \equiv \gamma^{1/2}\left(\frac{g_\star(\Tform)}{106.75}\right)^{1/4}\frac{\rho_{\rm BH}(\Tform)}{\rhotot(\Tform)},
\end{align}
where $g_{\star, \rm form}$ are the relativistic degrees of freedom at formation.
Having defined the main parameters that characterize the PBH population, we consider now its time evolution in the Early Universe.
Such time evolution will be governed by the Friedmann equation for the comoving energy densities $\varrho_i = a^{3(1+w_i)}\rho_i$, with $w_i$ the equation-of-state parameter for each component, $i={\rm \{rad, PBH\}}$~\cite{Barrow:1991dn, Gutierrez:2017ibk, Cheek:2021cfe}
\begin{subequations}
\begin{align}\label{eq:FBEqs}
 \frac{d\varrho_{\rm rad}}{dN_e}  &= -\frac{a}{H}\frac{d \ln M}{d t}\varrho_{\rm PBH},~ \\
  \frac{d\varrho_{\rm PBH}}{dN_e} & = \frac{1}{H}\frac{d \ln M}{d t}\varrho_{\rm PBH}\,,
\end{align} 
\end{subequations}
where we have chosen the number of e-folds, $N_e = \ln(a)$, where $a$ is the scale factor, as our time variable, and the Hubble parameter $H$ is
\begin{align}
 H^2=\frac{8\pi}{3 M_{\rm Pl}^2}(a^{-4}\varrho_{\rm SM} + a^{-3}\rho_{\rm PBH}) \,.
\end{align}
Given that the PBH population acts as a matter component, PBH domination may occur when~\cite{Barrow:1991dn,Hooper:2019gtx,Lunardini:2019zob}
\begin{align}\label{eq:beta_crit}
  \beta^\prime \gtrsim \beta^\prime_{c} \equiv 2.5\times 10^{-16} \left(\frac{g_\star(\Tform)}{106.75}\right)^{-1/4}\left(\frac{\MPBHini}{10^{10}~{\rm g}}\right)^{-1}.
\end{align}

Various constraints exist on the parameter space defined by $\beta^\prime$ and $\MPBHini$, depending on whether and when the PBH population evaporated throughout cosmological history~\cite{Carr:2020gox,Carr:2020xqk}.
PBHs having $\MPBHini\gtrsim 10^8~{\rm g}$ are expected to evaporate during the epochs of BBN and CMB formation.
Therefore, we can expect strong constraints on the initial PBH abundance for such masses~\cite{Carr:2020gox,Carr:2020xqk,Kohri:1999ex,Keith:2020jww,Acharya:2020jbv,Boccia:2024nly}.
Here, we will consider additional bounds coming from the production of neutrinos that contribute to the number of relativistic degrees of freedom, as constrained by CMB observations.

\section{Bounds from $N_{\rm eff}$}
\label{sec:PBHBoundsNeff}

Let us begin by reviewing the definition of $N_{\rm eff}$ in the context of standard cosmology. 
Since neutrino decoupling occurs shortly before electron-positron annihilation, the photon plasma undergoes heating, whereas the neutrino background remains largely unaffected~\cite{Dimastrogiovanni:2017tvd}. 
Consequently, the radiation energy density after annihilation can be expressed as
\begin{align}
    \rho_r = \frac{\pi^2}{30} \left(2 + \frac{7}{8}2N_{\nu}\left(\frac{T_\nu}{T}\right)^{4}\right) T^4,
\end{align}
where $\rho_r$ denotes the energy density of relativistic species in the universe, $N_\nu = 3$ represents the number of neutrino species in the SM, and $T$ ($T_\nu$) corresponds to the photon (neutrino) temperature.
More generally, the effective number of relativistic species, $N_{\rm eff}$, is defined as
\begin{equation}
    N_{\mathrm{eff}} = \left(\frac{11}{4}\right)^{4/3} \left(\frac{T_\nu}{T}\right)^{4} N_{\nu}.
    \label{eq:NeffNnu}
\end{equation}
Assuming there are no additional sources of neutrinos and considering instantaneous decoupling, the ratio of temperatures is given by $T_\nu/T = (4/11)^{1/3}$. 
However, taking into account the interactions between electrons and neutrinos during electron-positron annihilation, a more precise calculation yields the $\Lambda$CDM value of $N_{\rm eff}^{\rm \Lambda CDM} = 3.044$, see Refs.~\cite{Bennett:2020zkv, Froustey:2020mcq}.

Introducing PBHs into the analysis leads to two significant effects:
\begin{enumerate}

    \item {\bf Particle injection.}  The first effect arises from the fact that PBHs emit neutrinos, thereby increasing $N_{\nu}$ and, consequently, $N_{\mathrm{eff}}$. More formally, since these additional neutrinos contribute to the radiation energy density $\rho_r$, they introduce an additional term in Eq.~\eqref{eq:NeffNnu}. This contribution can be effectively absorbed into $N_{\mathrm{eff}}$, leading to its enhancement.

    \item {\bf Entropy injection.}  The second effect comes from PBHs emitting other SM particles, such as photons. If these particles are released after neutrino decoupling but before photon decoupling, their energy is injected into the photon plasma. Since, after neutrino decoupling, the photon plasma is no longer in thermal equilibrium with the neutrino background, this injection preferentially heats the photon plasma relative to the neutrino plasma. As a result, the temperature ratio $T_\nu/T$ decreases, leading to a reduction in $N_{\mathrm{eff}}$, as indicated by Eq.~\eqref{eq:NeffNnu}.
    
\end{enumerate}
The first effect was used in Refs.~\cite{Hooper:2019gtx,Lunardini:2019zob,Masina:2020xhk,Masina:2021zpu} to derive constraints on the PBH abundance, considering the effect of the emission of light particles having different spins. Specifically, Ref.~\cite{Lunardini:2019zob} considered the case of emission of right-handed neutrinos, after assuming neutrinos to be Dirac particles.
By employing Eq.~(3.8) from Ref.~\cite{Lunardini:2019zob}, we account for the first effect while neglecting the second. It is important to note that the second effect can result in $\Delta N_{\mathrm{eff}} < 0$, whereas Eq.~(3.8) imposes the constraint $\Delta N_{\mathrm{eff}} \geq 0$. This approximation remains valid before neutrino decoupling, as the additional emitted photons thermalize with the neutrinos in the plasma, ensuring that the ratio $T_\nu/T$ remains unaffected after decoupling. However, after neutrino decoupling, these emitted photons influence the temperature ratio and could lead to a decrease in $N_{\mathrm{eff}}$, meaning that their contribution can no longer be disregarded.

To determine the modification to $N_{\mathrm{eff}}$ due to PBH emission after neutrino decoupling, we consider the contribution of an additional dark radiation component, $\rho_{\rm DR}$, to the standard radiation energy density,
\begin{align}
    \rho_r &= \frac{\pi^2}{30} \left(2 + \frac{7}{8}2N_{\nu}\left(\frac{T_\nu}{T}\right)^{4}\right) T^4 + \rho_{DR} \notag\\
    &= \frac{\pi^2}{30} \left(2 + \frac{7}{8}2(N_{\nu}+N_{DR})\left(\frac{T_\nu}{T}\right)^{4}\right) T^4,
\end{align}
where $N_{\rm DR}$ is defined analogously to the standard neutrino contribution $N_\nu$ as
\begin{equation}
    N_{DR} = \frac{\rho_{DR}}{\rho_r^{\rm SM}}  \left(\frac{8}{7}\left(\frac{T_\nu}{T}\right)^{-4} + N_{\nu}\right).
    \label{NDRrhoDR}
\end{equation}
Here, $\rho_r^{\rm SM}$ represents the radiation energy density in the Standard Model when no dark radiation is present, i.e., $\rho_{DR} = 0$. This definition follows from the standard expression for $N_{\mathrm{eff}}$, as given in Eq.~\eqref{eq:NeffNnu}.

Thus, in the presence of a dark radiation component, the effective number of relativistic species is given by
\begin{align}
    N_{\mathrm{eff}} &= \left(\frac{11}{4}\right)^{4/3} \left(\frac{T_\nu}{T}\right)^{4} (N_{\nu}+N_{DR}).
    \label{NeffNnuNDR}
\end{align}
Substituting the definition of $N_{\rm DR}$ from Eq.~\eqref{NDRrhoDR}, we obtain the modified expression for $N_{\mathrm{eff}}$ that accounts for both additional relativistic degrees of freedom and changes in the radiation temperature:
\begin{equation}
    N_{\mathrm{eff}} = \left(\frac{11}{4}\right)^{4/3} \left(\frac{T_\nu}{T}\right)^{4} \left(N_{\nu} + \frac{\rho_{DR}}{\rho_r^{\rm SM}}  \left(\frac{8}{7}\left(\frac{T_\nu}{T}\right)^{-4} + N_{\nu}\right)\right).
\end{equation}
In this context, the fraction $\rho_{DR}/\rho_r^{\rm SM}$ must be evaluated at the moment when we seek to determine $N_{\mathrm{eff}}$. If this fraction is known at a different moment, we must translate it to the desired epoch. To first approximation, this ratio evolves primarily due to species annihilations occurring between the two moments, which introduces the usual \emph{effective number of relativistic species} scaling factors, see Ref.~\cite{Hooper:2019gtx,Lunardini:2019zob}. However, this assumption does not hold if there is energy injection between these moments.

In the cases of interest here, the available information corresponds to the epoch when PBHs have just evaporated, whereas the desired moment is the matter-radiation equality.
Since PBHs have already fully evaporated during this interval, no additional energy injection is expected, allowing us to apply the standard scaling behavior discussed in Ref.~\cite{Lunardini:2019zob}
\begin{equation}
    \frac{\rho_{DR}(T_{eq})}{\rho_r^{\rm SM}(T_{eq})} = \frac{\rho_{DR}(T_{ev})}{\rho_r^{\rm SM}(T_{ev})} \left(\frac{g_*(T_{ev})}{g_*(T_{eq})}\right) \left(\frac{g_{*S}(T_{eq})}{g_{*S}(T_{ev})}\right)^{4/3}.,
\end{equation}
where $T_{eq}\approx 0.8~{\rm eV}$ is the temperature at matter-radiation equality, and $g_{*S}(T)$ are the entropic relativistic degrees of freedom.
Substituting this relation, we obtain the final expression for $N_{\mathrm{eff}}$:
\begin{equation}
    N_{\mathrm{eff}} = \left(\frac{11}{4}\right)^{4/3} \left(\frac{T_\nu}{T}\right)^{4} \left(N_{\nu} + \frac{\rho_{DR}(T_{ev})}{\rho_r^{\rm SM}(T_{ev})} \left(\frac{g_*(T_{ev})}{g_*(T_{eq})}\right) \left(\frac{g_{*S}(T_{eq})}{g_{*S}(T_{ev})}\right)^{4/3}  \left(\frac{8}{7}\left(\frac{T_\nu}{T}\right)^{-4} + N_{\nu}\right)\right).
    \label{Neff_full_calc}
\end{equation}
The first term in this equation corresponds to ``effect 2'', arising from a modification in the temperature ratio, which may deviate from its standard value. In the second term, substituting $T_\nu/T \approx (4/11)^{1/3}$ recovers ``effect 1'', which accounts for the additional contribution of dark radiation emitted by PBHs. By refraining from making this substitution, we retain a correction term that slightly reduces the increase in $N_{\mathrm{eff}}$ due to ``effect 1''. This correction emerges from the interplay between both effects: if they had been treated separately and then summed, this correction would not appear.

Using Eq.~\eqref{Neff_full_calc}, we can compute $N_{\mathrm{eff}}$ for any given initial PBH mass $\MPBHini$ and abundance $\beta^\prime$. If the resulting value of $N_{\mathrm{eff}}$ is inconsistent with current observational constraints, we can exclude that particular $\beta'$ at the given PBH mass. This enables us to place mass-dependent bounds on $\beta'$ based on this effect. Importantly, since the modification to $N_{\mathrm{eff}}$ can be either positive or negative, both upper and lower observational limits on $N_{\mathrm{eff}}$ must be taken into account.

To constrain the effects of PBH evaporation on $N_{\mathrm{eff}}$, we consider the observational bounds that were recently reported by ACT \cite{ACT:2025tim}. More specifically, we consider the bound on eq. 47 of that paper, $N_\mathrm{eff} = 2.86 \pm 0.14$. This bound uses data from observations of the CMB and CMB lensing from Planck \cite{Planck:2019nip, Planck_2018_Cosmological_Paremeters_2020A&A...641A...6P, Planck:2018lbu, Carron:2022eyg} and ACT \cite{ACT:2025fju, ACT:2023kun}, BAO observations from DESI \cite{DESI:2024mwx}, and a compilation of observations of the primordial helium abundance recommended by the PDG \cite{ParticleDataGroup:2024cfk}. However, the bound does not use any information from BBN theory, and instead treats the primordial helium fraction as a free parameter that is constrained by observations. This is exactly what we want here, since we do not expect standard BBN theory to be valid here. This is because, apart from changing $N_\mathrm{eff}$, light PBHs have other effects on BBN that we are not taking into account here. Besides, most of the PBHs in the mass range that we are considering evaporate after BBN, and thus only significantly alter $N_\mathrm{eff}$ after BBN. 

Thus, at the $2\sigma$ confidence level, we have:
\begin{equation}
    2.58 < N_{\mathrm{eff}} < 3.14.
\end{equation}

We note that this bound assumes a $\Lambda \mathrm{CDM} + N_\mathrm{eff} + Y_\mathrm{He}$ model, and could potentially be loosened in a more general model with more free parameters.

Since these constraints apply to $N_{\mathrm{eff}}$ at the time of CMB emission, we consider only PBHs that have fully evaporated before this epoch. This ensures that their complete contribution to $N_{\mathrm{eff}}$ is present by the time the CMB was emitted. Consequently, we restrict our analysis to initial PBH masses up to approximately $2 \times 10^{13}$ g.

To quantify the neutrino emission from PBHs, we analyze both Dirac and Majorana neutrino cases, as the number of emitted neutrinos differs between these scenarios. 
To account for changes in the neutrino-to-photon temperature ratio due to the emission of other particles, we use the fact that the photon energy density scales as $\rho_\gamma \propto T^4$. 
This allows us to write
\begin{equation}
    \left(\frac{T_\nu}{T}\right)^4 = \left(\frac{T_\nu}{T}\right)^4_{\rm standard} \cdot \left(\frac{\rho_{\gamma}^{\rm SM}}{\rho_{\gamma}^{\rm SM}+ \rho_{\gamma}^{\rm  PBH}}\right),
\end{equation}
where $\rho_{\gamma}^{\rm SM}$ is the standard photon energy density and $\rho_{\gamma}^{\rm PBH}$ represents the energy density of photons emitted by PBHs. The standard temperature ratio is given by:
\begin{equation}
    \left(\frac{T_\nu}{T}\right)^4_{\rm standard} = \left(\frac{4}{11}\right)^{4/3} \frac{N_{\rm eff}^{\rm\Lambda CDM}}{3}.
\end{equation}
For full accuracy, $\rho_{\gamma}^{\rm SM}$ should also include photons emitted before neutrino decoupling, as these remain present but do not affect the temperature ratio. This is a small correction but is accounted for in our analysis.
The standard photon energy density is given by
\begin{equation}
    \rho_{\gamma}^{\rm SM} = \frac{\pi^2}{15} T(\beta' = 0)^4,
\end{equation}
where $T(\beta' = 0)$ denotes the plasma temperature in the absence of PBHs.

An important consideration is that neutrino decoupling occurs approximately one second after the Big Bang. Prior to this, neutrinos and photons remain in thermal equilibrium due to weak interactions. Consequently, any neutrinos or photons emitted before this epoch will thermalize and will not alter the neutrino-photon temperature ratio or their respective energy densities. Therefore, neutrinos and photons emitted before neutrino decoupling can generally be ignored.
However, there is a crucial exception: for Dirac neutrinos, half of the emitted primary neutrinos are right-handed, meaning they do not participate in weak interactions and thus do not thermalize. As a result, we include these right-handed neutrinos even if they were emitted before neutrino decoupling.

This consideration significantly impacts the PBH mass range for which our analysis is relevant. The primary effects that we include are the contribution of all emitted neutrinos and the reduction in $N_{\mathrm{eff}}$ due to additional particle emission. These effects are irrelevant for PBHs that evaporate before neutrino decoupling, as all additional emitted particles thermalize and do not affect $N_{\mathrm{eff}}$. Thus, our refined approach is particularly relevant for PBHs evaporating after neutrino decoupling, corresponding to PBH masses greater than $\sim 10^9$ g. Consequently, we focus our analysis on PBHs with masses above $10^9$ g. For lower masses, if neutrinos are Dirac particles, the constraints derived in Ref.~\cite{Lunardini:2019zob} remain applicable. If neutrinos are Majorana particles, however, no meaningful bounds on $N_{\mathrm{eff}}$ can be placed at these lower masses.

Another potential effect that could influence our analysis arises from the formation of hot spots around PBHs, as discussed in Refs.~\cite{He:2022wwy,He:2024wvt}. In the early universe, these localized regions of increased temperature could modify our reasoning if neutrinos thermalized before escaping the hot spot. This effect is most relevant at very early times, corresponding to extremely high plasma temperatures.
However, we have verified that the plasma temperature at which neutrinos cease to thermalize within the hot spot is significantly higher than the temperature at which neutrinos decouple from the plasma. As a result, these hot spots would only impact neutrinos that we were already disregarding in our analysis. Therefore, it is safe to neglect this effect.

\subsection{Effect of the emitted particles' interactions}
\label{subsec:interactions}

After being emitted, particles will still interact with the plasma, which could impact their effect on $N_\mathrm{eff}$. There are three relevant cases here: stable particles that feel the electromagnetic (and potentially the strong) interaction, such as photons, electrons or protons, stable particles that only feel the weak interactions, i.~e., neutrinos, and unstable particles.

For stable particles that feel the electromagnetic interaction, such as photons or electrons, we expect that they will thermalize via EM scatterings, thus injecting all their energy into the photon plasma and raising its temperature, thereby reducing $N_\mathrm{eff}$, as explained in the previous subsection. However, one might ask if such scatterings could occur via the weak interactions, leading to the injection of additional neutrinos. This can happen if the center-of-mass energy of these interactions fulfills $\sqrt{s} \gtrsim 10 \mathrm{GeV}$. Otherwise, weak interactions will be negligible compared to electromagnetic and strong interactions. To assess the relevance of such scatterings, we first compute the center-of-mass energy of these collisions:  
\begin{align}  
    \sqrt{s} &\sim \sqrt{2\, E_p\, E_{\rm BH}} \sim \sqrt{12\, T\, T_{\rm BH}},\notag\\  
    &\sim 10~{\rm GeV} \left(\frac{T}{\rm 1~MeV}\right)^{1/2}\left(\frac{\rm 10^9~g}{M}\right)^{1/2},  
\end{align}  
where we assume that a plasma particle has an average energy $E_p \sim T$, while a particle produced via PBH evaporation has $E_{\rm BH} \sim 6\, T_{\rm BH}$~\cite{Cheek:2021odj}, and we neglect particle masses.  
This result implies that for a plasma temperature of $T \lesssim 1$ MeV, PBHs with instantaneous masses below $M \lesssim 10^7~{\rm g}$ could produce sufficiently energetic particles to interact via weak interactions, leading to neutrino production. However, since we consider PBHs with initial masses larger than $10^9$ g, the fraction of particles emitted with such high energies when the PBH mass decreases below $10^7$ g is small relative to the total number of emitted particles.  
We estimate this suppression by computing the ratio $R$ of the total number of particles emitted while the PBH mass is in the range $M \in [10^7, 10^9]~{\rm g}$ to those emitted when the mass is in the interval $M \in [M_P, 10^7]~{\rm g}$,  
\begin{align}  
    R = \frac{N_i[M_P, 10^7~{\rm g}]}{N_i[10^7~{\rm g}, 10^9~{\rm g}]} \approx 0.01\%,  
\end{align}  
where  
\begin{align}  
    N_i[M_{\rm min}, M_{\rm max}] =\int_{M_{\rm min}}^{M_{\rm max}} dM \left(\frac{dM}{dt}\right)^{-1}\,\int dp\, \frac{d^{2}N_i}{dt\, dE}.  
\end{align}  
For PBHs with larger initial masses evaporating in a colder Universe, this ratio decreases even further, allowing us to safely neglect the contribution of such weak scatterings with the plasma in neutrino production. Hence, we can safely assume that these particles always thermalize electromagnetically\footnote{Or via the strong force, but this makes no difference because baryons and photons always remain in thermal equilibrium before recombination, and we are always considering times before recombination here.}, thus injecting their energy into the photon plasma.

For the case of neutrinos, they do not feel the electromagnetic and strong interactions, so the previous reasoning does not hold. Indeed, active left-handed neutrinos emitted from PBHs tend to have higher energies than those in the plasma, so their interaction probability is increased, allowing them to remain coupled to the plasma even after neutrino decoupling. We incorporate this effect in an energy-dependent manner by defining an energy-dependent ``decoupling'' temperature, so high-energy neutrinos interact until the temperature of the universe falls below their corresponding decoupling temperature. More details on this approach are provided in Appendix~\ref{ap:EnergyDependentNeutrinoDecoupling}.

Let us now discuss the case of unstable particles. Unstable particles that can decay via the electromagnetic or the strong force can be expected to do so, and will not produce any neutrinos, so their energy will ultimately contribute to the photon plasma. Note that, even if they interact electromagnetically before decaying, this will not change the fact that their energy will go to the photon plasma, so it is irrelevant for us. However, what happens with emitted particles that are unstable and decay via weak interactions, such as muons, pions and kaons? If they decay before interacting, these particles will significantly contribute to the secondary neutrino spectrum. However, if they interact before they decay, via the electromagnetic or the strong interaction (Weak scatterings are, as explained above, negligible compared to EM or strong scatterings), this will cause them to inject energy to the photon plasma and, since they will have less energy once they decay, it will reduce the amount of energy they inject into the neutrino plasma when they decay. Both of these effects will reduce $N_\mathrm{eff}$. Therefore, determining whether such particles decay before thermalizing or thermalize before decaying will be very important for our bounds. To calculate this, we (approximately) calculate the scattering rate $\Gamma_{\mu, \mathrm{scatter}}$ of these particles in the plasma and compare it to their decay rate $\Gamma_{\mu,\mathrm{decay, lab}}$, as explained in more detail in appendix \ref{ap:MuonsPionsKaons}.

If the scattering rate is much larger than the decay rate, then muons, pions and kaons will thermalize before decaying, thereby injecting most of their energy into the photon plasma and reducing $N_\mathrm{eff}$. The amount of energy they inject into the neutrino plasma will be lower than the energy they have left when they decay, which will be of the order of their mass, which is typically much smaller than their energy at emission, of the order of the Hawking temperature. Therefore, we can approximate that, in this regime, all the energy of these particles will go to the photon plasma and reduce $N_\mathrm{eff}$. For the case of pions\footnote{The limit would be slightly lower for the case of kaons, but, since pions make up a considerably larger part of the emitted spectrum, we neglect this and use pions. For muons, the limit would be less strict anyway.}, considering that the ratio of rates $\Gamma_{\mu, \mathrm{scatter}} / \Gamma_{\mu,\mathrm{decay, lab}}$ is ''much larger than 1'' if it is at least 10, we get that this regime applies if $M < 7.0 \cdot 10^{10} \mathrm{g}$.

Conversely, if the scattering rate is much smaller than the decay rate, then muons, pions and kaons will decay before thermalizing. Then, by knowing the decay properties of these particles, one can calculate their contribution to the secondary spectra of neutrinos, which will raise $N_\mathrm{eff}$, as well as their contribution to the secondary spectra of other particles, such as photons or electrons, which will lower $N_\mathrm{eff}$. We calculate these secondary spectra by using the numerical code {\tt BlackHawk}~\cite{Arbey_2019_BlackHawk, Arbey:2021ysg}, as described in more detail in the next section. For the case of muons, considering that the ratio of rates is ``much smaller than 1'' if it is at most 0.1, we get that this regime applies if $M > 3.8 \cdot 10^{11} \mathrm{g}$. There is also a region where muons already start thermalizing, but pions and kaons (whose lifetime is considerably shorter) still decay without thermalizing. Since, before pion decay, the energy emitted in form of pions is generally considerably larger than the energy emitted in form of muons, in this regime, assuming that all these particles decay without thermalizing should still yield approximately correct results. This ``partially valid'' regime applies for $1.6 \cdot 10^{11} \mathrm{g} < M < 3.8 \cdot 10^{11} \mathrm{g}$.

There is also an intermediate regime ($7.0 \cdot 10^{10} \mathrm{g} < M < 1.6 \cdot 10^{11} \mathrm{g}$) where scattering rates and decay rates are of the same order. A description of this regime would require a more detailed treatment of the involved scattering reactions, which is beyond the scope of this work. Therefore, we do not derive bounds in that regime.

\subsection{Numerical calculation of the bounds}

To numerically compute the bounds on $N_{\mathrm{eff}}$, we first determined the spectra of neutrinos emitted by PBHs across a range of masses. For this purpose, we utilized the code {\tt BlackHawk}~\cite{Arbey_2019_BlackHawk, Arbey:2021ysg}, which calculates the Hawking radiation spectra of black holes. 
The computed spectra include both \emph{primary} neutrinos, which are directly emitted from the PBH, and \emph{secondary} neutrinos, which originate from the decays of other particles emitted by the PBH. 
In addition to neutrinos, we also used {\tt BlackHawk} to compute the spectra of other emitted particles, particularly photons and electrons, which contribute to the heating of the plasma. This information is crucial for accurately determining the neutrino-photon temperature ratio, a key quantity required in the calculation of $N_{\mathrm{eff}}$ using Eq.~\eqref{Neff_full_calc}. 
By including both primary and secondary neutrino spectra, as well as the thermal effects of other emitted particles, we ensure a precise evaluation of the impact of PBH evaporation on $N_{\mathrm{eff}}$.

To obtain the secondary spectra, {\tt BlackHawk} relies on precomputed tables generated with the particle physics code {\tt PYTHIA}~\cite{Sj_strand_2015_Pythia}. In the regime where pions, muons and kaons are expected to decay, we use the option to use the tables that do not consider them as final particles and describe their decay. In the regime where they are expected to thermalize, we use the tables that consider them as final particles, so that we know their spectrum and can add their energy to the photon energy density later.

To relate the PBH abundance $\beta'$ to the emitted neutrino spectra for a given mass, we first calibrated the spectra following the procedure outlined in Appendix~\ref{ap:SpectraCalibration}. 
Once calibrated, the spectra were integrated over time to obtain the total neutrino spectrum, accounting for redshifting due to cosmic expansion. The time-dependent scale factor was extracted using the cosmology code \texttt{CLASS}~\cite{Diego_Blas_2011_CLASS}, which was also employed for cross-checks. We then integrated the total spectra over energy to compute the total energy density of emitted neutrinos. The temperatures and radiation energy densities at PBH evaporation, as functions of PBH mass and $\beta'$, were obtained using tables generated with \texttt{FRISBHEE}~\cite{Cheek:2022mmy, Cheek:2022dbx, Cheek:2021odj}.

Our results were derived using two approaches: one relying on the \texttt{FRISBHEE} tables (with \texttt{CLASS} used only for the scale factor), and another based exclusively on \texttt{CLASS} outputs. The latter is expected to be less accurate before electron-positron annihilation but serves as a useful consistency check. Since \texttt{CLASS} assumes standard cosmology, our bounds are valid only in the absence of a PBH-dominated phase. However, as shown in Ref.~\cite{Lunardini:2019zob}, existing constraints already exclude PBH domination in the relevant mass range, ensuring the reliability of our results.

We followed a similar procedure using primary and secondary photon spectra from \texttt{BlackHawk} to determine the change in the neutrino-photon temperature ratio. We integrated them over time to obtain the total spectrum and then over energy to compute the injected energy density. In addition to photons, we accounted for electrons and positrons, assuming they thermalize and transfer their energy to the photon plasma. In the regime where they thermalize, muons, charged pions and charged kaons are also accounted for in this way. The temperature change was then derived using $\rho_{\gamma} \propto T^4$.
With the neutrino energy density and the modified neutrino-photon temperature ratio as functions of $\beta'$, we applied Eq.~\eqref{Neff_full_calc} to compute $N_{\mathrm{eff}}$. By identifying the values of $\beta'$ that exceed observational constraints, we established exclusion bounds on $\beta'$ for different PBH masses.

To assess the impact of various physical effects, we computed bounds with and without secondary Hawking radiation, as well as with and without the temperature ratio modification. Since the Dirac/Majorana nature of neutrinos remains unknown, we considered both cases.
For PBHs of mass $M=10^{12}$ g, using the \texttt{CLASS}-only method, we also examined scenarios including graviton emission and Kerr PBHs, both with and without graviton emission. Since gravitons, like neutrinos, are relativistic and weakly interacting, they contribute to $N_{\mathrm{eff}}$ in the same manner as right-handed neutrinos and are not excluded if emitted before neutrino decoupling.

\section{Results}\label{sec:Results}


In this section, we present our main results: new constraints on the abundance of Schwarzschild PBHs in the mass range from $10^9$ g to approximately $2 \times 10^{13}$ g. These constraints arise from incorporating physical effects that have not been previously considered in the literature: the contribution of secondary neutrino emission and the modification of the neutrino-photon temperature ratio due to the entropy injection from Hawking radiation beyond neutrinos.  

\begin{figure}
    \centering
    \includegraphics[width=\linewidth]{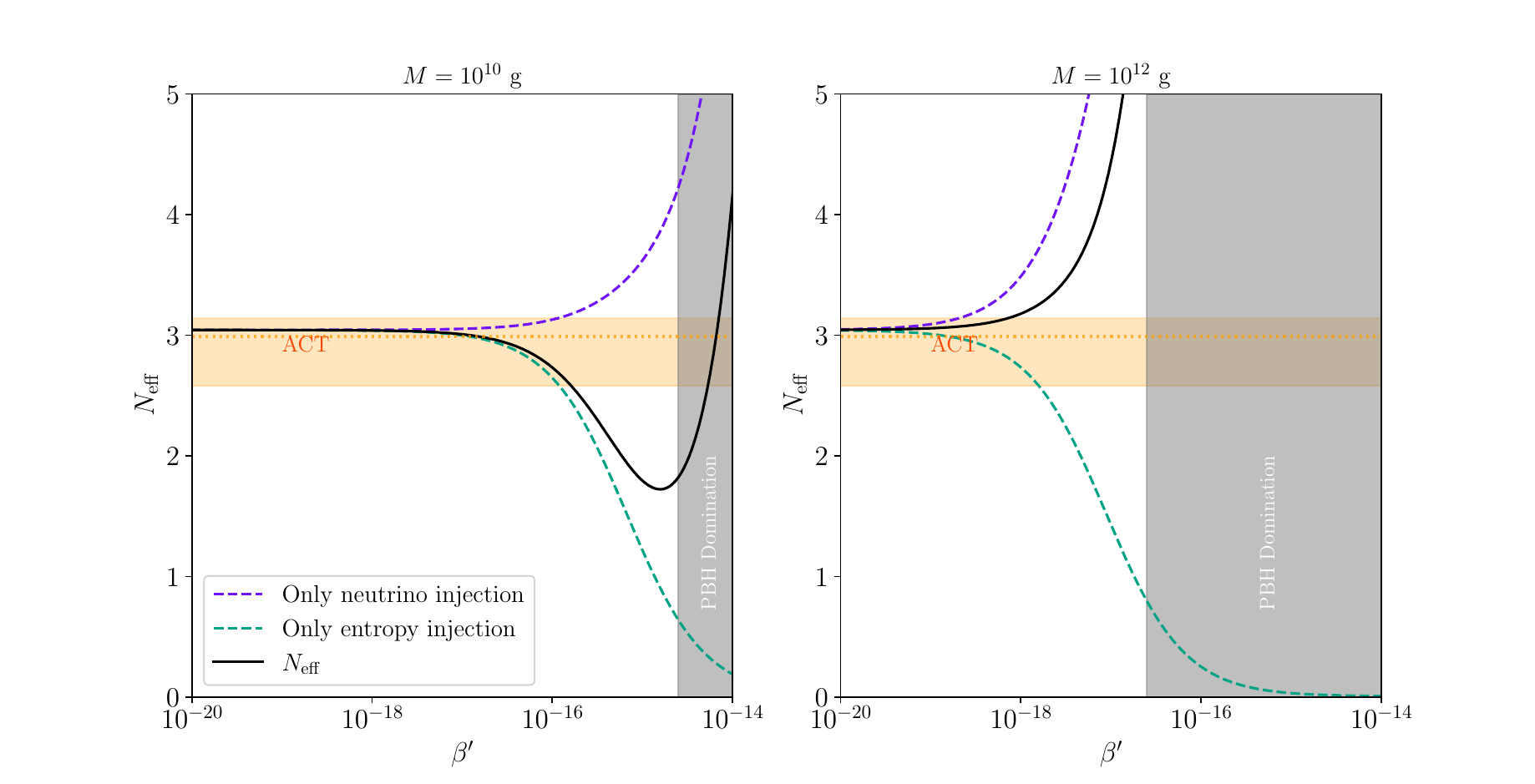}
    \caption{Effects of Hawking radiation from Schwarzschild PBHs with masses $M=10^{10}$ g (left) and $M = 10^{12}$ g (right) on $N_{\mathrm{eff}}$, assuming Majorana neutrinos, as a function of PBH abundance $\beta'$.  The contribution from neutrino emission (purple dashed line) increases $N_{\mathrm{eff}}$, while entropy injection reducing the neutrino-photon temperature (green dashed line) lowers it.  The total effect (black line) combines both contributions. Observational $2\sigma$ bounds from Ref.~\cite{ACT:2025tim} are shown as an orange band. The gray region represents the part of parameter space where PBH domination occurs, so our method is no longer valid.}
    \label{NeffEffectsPlot}
\end{figure}

We first examine the individual impact of two effects mentioned before.
Fig.~\ref{NeffEffectsPlot} illustrates how the PBH abundance $\beta'$ affects $N_{\mathrm{eff}}$ for Schwarzschild PBHs with masses $M_1= 10^{10}$ g and $M_2 = 10^{12}$ g, assuming Majorana neutrinos. Note that $M_1$ lies in the low-mass regime where muons, pions and kaons thermalize, whereas $M_2$ lies in the high-mass regime where they decay without thermalizing. While the Dirac case would yield slightly different quantitative results, the qualitative behavior remains unchanged, making the following discussion applicable to both cases.  
The figure separately presents the two competing effects: (i) an increase in $N_{\mathrm{eff}}$ due to neutrino emission as Hawking radiation (purple dashed) ---labeled as \emph{only neutrino injection}--- and (ii) a decrease in $N_{\mathrm{eff}}$ caused by plasma heating from photons and electrons (green dashed) ---labeled as \emph{only entropy injection}----, which reduces the neutrino-photon temperature ratio $T_\nu / T$. The net change in $N_{\mathrm{eff}}$, shown in black, is the sum of these effects.  
The observational $2\sigma$ limits from Ref. \cite{ACT:2025tim} are included in the figure as the orange band. 
The bound on $\beta'$ is determined by the point where $N_{\mathrm{eff}}$ exits the observational limits — values of $\beta'$ outside this band are excluded.

In the high-mass regime, $N_\mathrm{eff}$ increases, as the change is dominated by the neutrino injection effect, and the bound comes from the point where $N_\mathrm{eff}$ exceeds its upper limit. Notably, the plasma heating effect delays this crossing, weakening the constraints compared to a scenario without it.  
Thus, Fig.~\ref{NeffEffectsPlot} demonstrates that plasma heating allows for a higher neutrino contribution to $N_{\mathrm{eff}}$ without violating observational bounds, effectively masking additional neutrinos. 

In the low-mass regime, where muons, pions and kaons essentially introduce no energy into the neutrino plasma and instead inject all their energy into the photon plasma, the entropy injection effect is stronger, so $N_\mathrm{eff}$ decreases, and the bound is given by the point where it crosses its lower limit, which is delayed by the competing neutrino injection effect. Note that, in this regime, an intermediate value of $\beta'$ that is within the bound is a way to reduce $N_\mathrm{eff}$, providing a potential explanation for ACT's mild preference for $N_\mathrm{eff} < 3$ \cite{ACT:2025tim}.

One can also see that, at high values of $\beta'$, $N_\mathrm{eff}$ re-enters the observational band, which would in principle lead to an additional window of allowed $\beta'$ values. However, this happens in the regime of PBH domination, where we do not expect our method to be valid, so we cannot confirm the existence or the location of that window. As we will see later, this is fine, because BBN bounds \cite{Carr:2020gox} exclude the PBH domination region anyway.

We now compare the resulting bounds for Dirac and Majorana neutrinos and examine the impact of including secondary neutrinos. 

\begin{figure}
    \centering
    \includegraphics[width=0.65\linewidth]{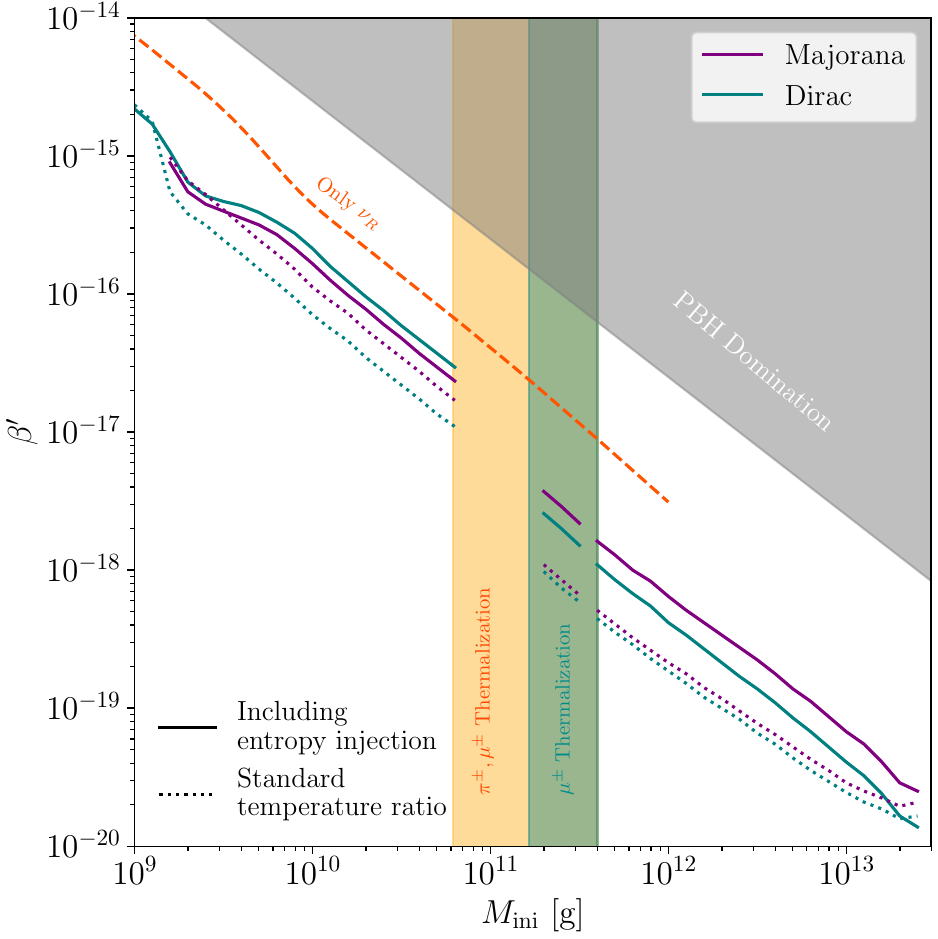}
    \caption{$2\sigma$ bounds on the abundance of Schwarzschild PBHs of various masses for Majorana (purple) and Dirac (teal) neutrinos, including entropy injection from PBH evaporation (full lines) and neglecting it (dotted lines). Bounds from \cite{Lunardini:2019zob} (including only right-handed neutrinos) are shown for comparison (orange dashed line). The green band shows the intermediate regime where muons partially thermalize before decaying and our bounds are only approximately valid. The yellow band shows the intermediate regime where pions partially thermalize before decaying and we report no bounds. To the bands' left, muons, pions and kaons fully thermalize before decaying. To the bands' right, they decay without thermalizing. The gray region represents the part of parameter space where PBH domination occurs, so our method is no longer valid.}
    \label{fig:TratioBounds}
\end{figure}

\begin{figure}
    \centering
    \includegraphics[width=0.65\linewidth]{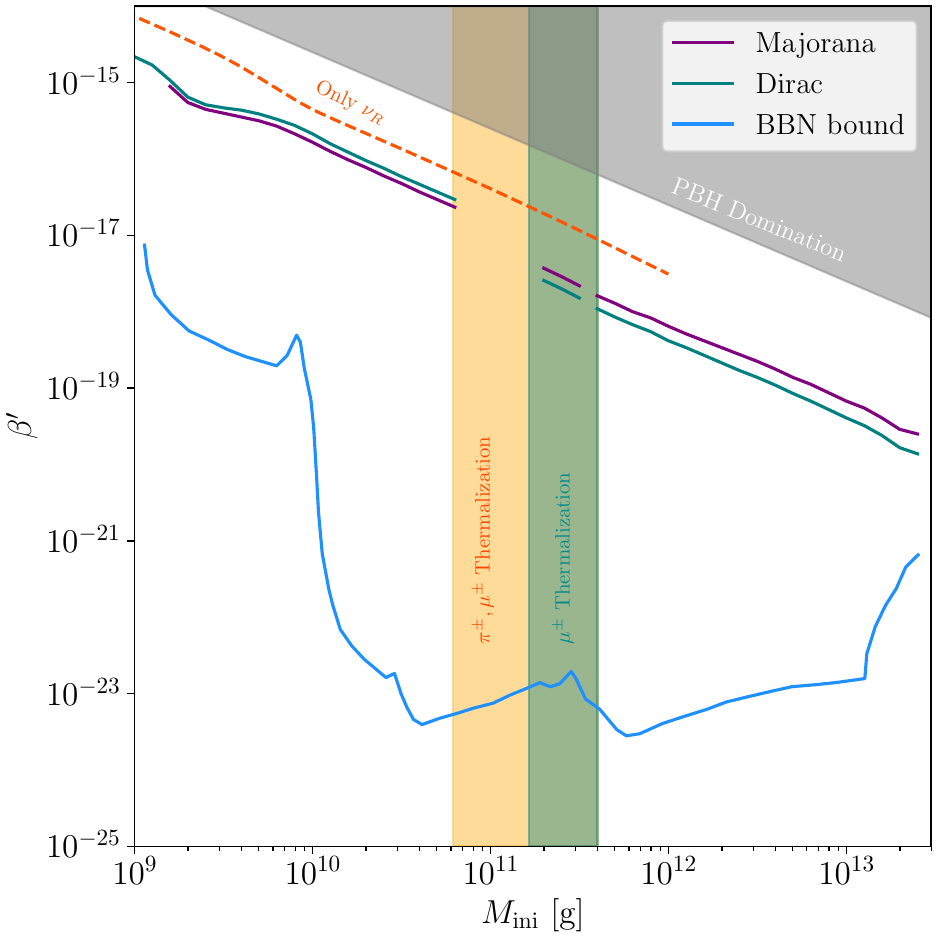}
    \caption{Bounds on the abundance of Schwarzschild PBHs of various masses for Majorana (purple) and Dirac (teal) neutrinos coming from $N_{\rm eff}$ measurements. For comparison, we show bounds from the same phenomenon that include only right-handed neutrinos, taken from \cite{Lunardini:2019zob} (orange dashed line), as well as bounds coming from Big Bang Nucleosynthesis, taken from \cite{Carr:2020gox} (blue line). The green band shows the intermediate regime where muons partially thermalize before decaying and our bounds are only approximately valid. The yellow band shows the intermediate regime where pions partially thermalize before decaying and we report no bounds. To the bands' left, muons, pions and kaons fully thermalize before decaying. To the bands' right, they decay without thermalizing. The gray region represents the part of parameter space where PBH domination occurs, so our method is no longer valid. All bounds have a confidence level of $2\sigma$.}
    \label{Neff_vs_BBN_Bounds}
\end{figure}

As previously mentioned, we computed our bounds using two methods: one primarily relying on {\tt FRISBHEE} with {\tt CLASS} providing the time-dependent scale factor, and another using {\tt CLASS} alone. 
Both methods show excellent agreement at intermediate masses. At low masses, discrepancies arise, as expected, since the {\tt CLASS} method is less precise in this range due to its treatment of electron-positron annihilation, making {\tt FRISBHEE} the more reliable approach. At high masses, deviations occur due to matter domination—PBH evaporation begins near or after the onset of matter domination, whereas {\tt FRISBHEE} assumes radiation domination. Consequently, at high masses, the {\tt CLASS} method is the more accurate one. Hence, for the bounds presented on Figs. \ref{fig:TratioBounds} and \ref{Neff_vs_BBN_Bounds}, we select the most reliable method for each mass range, {\tt FRISBHEE} for $M < 10^{12}$ g and {\tt CLASS} for higher masses.

Figure~\ref{fig:TratioBounds} presents the constraints on the initial PBH abundance $\beta^\prime$ as a function of the initial mass $M_{\rm ini}$ for Dirac and Majorana neutrinos. 
The dotted line represents the limits that would be obtained if the modification of the neutrino-photon temperature ratio due to particle injection were not included.  
As photon and electron emission heats the plasma, the neutrino-photon temperature ratio decreases, leading to a reduction in $N_{\mathrm{eff}}$. This effect competes with the increase in $N_{\mathrm{eff}}$ from neutrino Hawking radiation, ultimately weakening the bounds on $\beta'$. The figure also shows bands that indicate the intermediate regimes where muons and pions partially thermalize before decaying, and thus separate the low-mass regime where they fully thermalize before decaying from the high-mass regime where they decay without thermalizing.

For comparison, the figure also shows bounds from ref. \cite{Lunardini:2019zob}, which assume Dirac neutrinos, but only consider the increase in $N_{\mathrm{eff}}$ due to the emission of right-handed neutrinos, so they do not include the effect of the entropy injection or secondary neutrinos. If we switch off these effects in our calculations, we find good agreement with their results. This comparison shows us that the emission of secondary neutrinos coming from the weak decay of unstable particles significantly strengthens the bounds and reduces the difference between the Dirac and Majorana cases. This is expected, as secondary neutrinos originate from weak decays and are exclusively left-handed, making their abundance independent of whether neutrinos are Dirac or Majorana. Since secondary neutrinos dominate the total neutrino emission at these masses, the relative contribution of right-handed neutrinos in the Dirac case becomes less significant.


The temperature ratio effect weakens the constraints on $\beta'$, as expected. Note, however, that it does not fully compensate for the strengthening of the bounds due to the effect of secondary neutrinos, so taking both effects into account makes the constraints stricter than ignoring both of them. In the high-mass regime, the effect also increases the difference between the Dirac and Majorana cases. This is intuitive: since the reduction in $N_{\mathrm{eff}}$ partially offsets the increase from emitted neutrinos, the relative difference in neutrino emission between Dirac and Majorana cases becomes more pronounced. In the low-mass regime, the opposite is true, which also makes sense: In that regime, when including entropy injection, the bounds mainly come from the entropy injection effect, as the reduction in $N_\mathrm{eff}$ due to that effect is larger than its increase due to the neutrino injection, which becomes a relatively minor correction. Since the entropy injection is independent from the Dirac / Majorana nature of neutrinos, this reduces the difference between both cases.

Also note that, for the bounds in the low-mass regime including entropy injection, the bounds for Majorana neutrinos are stricter than those for Dirac neutrinos, whereas the Dirac bounds are stricter than the Majorana bounds in all other cases. This is because, as seen in fig. \ref{NeffEffectsPlot}, in the low-mass regime accounting for entropy injection, $N_\mathrm{eff}$ is reduced, and its bound comes from its observational lower limit, whereas the bound comes from the observational upper limit in all other cases. Since neutrinos being Dirac particles increases the neutrino injection, it raises $N_\mathrm{eff}$, thus strengthening the bounds in cases where they come from the upper limit on $N_\mathrm{eff}$, and weakening them when they come from the lower limit.

Figure~\ref{Neff_vs_BBN_Bounds} presents our final bounds, and also compares them with previous bounds from \cite{Lunardini:2019zob} and with BBN bounds from Ref.~\cite{Carr:2020gox}.
Even though our constraints are stricter than previous constraints coming from the same phenomenon, we find that they are less stringent than the BBN bounds, and thus do not exclude any previously allowed region of the PBH parameter space.

We now briefly analyze Kerr PBHs, characterized by the spin parameter $a_* = J/GM^2$, where $J$ is the black hole angular momentum, ranging from 0 (non-rotating) to 1 (maximally rotating). 
We computed $\beta^\prime$ bounds for $M = 10^{12}$ g, considering cases with and without graviton emission. 
Graviton emission, expected to be small for non-rotating PBHs but enhanced at high $a_*$, should strengthen the bounds by contributing to $N_{\mathrm{eff}}$, though its effect remains negligible except at high $a_*$, where it slightly lowers the constraints. 
Overall, while rotation affects the bounds, it at most weakens them by a factor of $\sim 2$, and since gravitons remain negligible in realistic cases, the bounds obtained before hold for rotating PBHs with at most an order $\sim 2$ correction.

\section{Conclusions}\label{sec:Conclusions}

The discovery of primordial black holes (PBHs) would provide a crucial test of their predicted thermal emission. Moreover, a significant population of PBHs evaporating in the early universe could drastically alter cosmological evolution, leading to deviations from standard predictions. This has motivated strong constraints on PBH masses, particularly for those evaporating during Big Bang Nucleosynthesis (BBN) or impacting Cosmic Microwave Background (CMB) observables.  

An additional key cosmological probe is the effective number of relativistic species, $N_{\rm eff}$, which has been measured through both BBN and CMB analyses. This observable is influenced by neutrino injection after decoupling and modifications to the neutrino-photon temperature ratio. In this paper, we have examined how a PBH population affects $N_{\rm eff}$ through both of these mechanisms.

We have shown that incorporating two additional effects strengthens the bounds on PBH abundance in the mass range $10^9$ g to $10^{13}$ g derived from $N_{\mathrm{eff}}$. 
Including neutrino emission from secondary Hawking radiation significantly tightens the constraints, while accounting for the heating of the photon plasma by other emitted particles reduces $N_{\mathrm{eff}}$ and weakens the bounds. Despite this partial cancellation, our results still yield somewhat stronger limits than those in Ref.~\cite{Lunardini:2019zob}.
However, our constraints are not the most stringent in this mass range. As shown in Fig.~\ref{Neff_vs_BBN_Bounds}, BBN bounds remain considerably stricter, meaning our results do not exclude any previously allowed regions in the PBH parameter space.  

The key takeaway from this study is not just the refined bounds but the underlying physics. We have demonstrated that reducing the neutrino-photon temperature ratio can lower $N_{\mathrm{eff}}$, allowing additional neutrinos (or other dark radiation components) to be effectively ``hidden'' without conflicting with observations. This effect could influence constraints on other phenomena, such as dark matter annihilation, where the production of photons or neutrinos depends on branching ratios. 
Additionally, analyzing spectral distortions in the CMB arising from the interaction of the PBH-emitted neutrinos with the neutrino background may improve the constraints derived here.
Exploring these broader implications, however, is beyond the scope of this work and is left for future studies.

\section*{Acknowledgements}

 GB and HS are supported by the Spanish grants  CIPROM/2021/054 (Generalitat Valenciana),  PID2023-151418NB-I00 funded by MCIU/AEI/10.13039/501100011033/, and by the European ITN project HIDDeN (H2020-MSCA-ITN-2019/860881-HIDDeN). HS is also supported by the grant FPU23/00257, MCIU.
YFPG has been supported by the Consolidaci\'on Investigadora grant CNS2023-144536 from the Spanish Ministerio de Ciencia e Innovaci\'on (MCIN) and by the Spanish Research Agency (Agencia Estatal de Investigaci\'on) through the grant IFT Centro de Excelencia Severo Ochoa No CEX2020-001007-S.
\appendixtitleon
\appendixtitletocon
\begin{appendices}
\section{Energy-dependent neutrino interactions}\label{ap:EnergyDependentNeutrinoDecoupling}

For our calculation of $N_{\mathrm{eff}}$, we only consider neutrinos that remain non-interacting after emission, as they contribute to increasing $N_{\mathrm{eff}}$. Neutrinos that interact after being emitted can transfer energy to both the neutrino and photon plasma. Before neutrino decoupling, neutrinos and photons remain in thermal equilibrium, so emitted particles do not alter $N_\mathrm{eff}$ and we can safely exclude them.

For emitted neutrinos that are emitted after plasma neutrinos have decoupled, but still interact after being emitted, their effect on $N_\mathrm{eff}$ is less clear. To calculate it, one would need to consider all the relevant reactions and whether their products contribute to the neutrinos, the photons or both, but such a procedure is beyond the scope of this work. Therefore, we always treat them with the approach that leads to more conservative bounds. In the high-mass regime where pions, muons and kaons decay without thermalizing and the bound on $N_\mathrm{eff}$ comes from its upper limit, we assume that these neutrinos contribute to both photons and neutrinos in a way that does not alter $N_\mathrm{eff}$, so we exclude them. In the low-mass regime where pions, muons and kaons thermalize before decaying and the bound on $N_\mathrm{eff}$ comes from its lower limit (so excluding these neutrinos would lead to a stricter bound), we instead include them and assume that they all contribute to the neutrino plasma.

In this appendix, we outline the procedure used to determine which neutrinos in the high-mass regime are excluded due to post-emission interactions. This applies only to left-handed neutrinos, as right-handed neutrinos (if they exist in the Dirac case) do not participate in weak interactions and thus do not thermalize.  

Since PBHs emit neutrinos with a spectrum of energies, higher-energy neutrinos have a greater likelihood of interacting. To account for this, we calculate the critical plasma temperature required for a neutrino of a given energy to interact. If a neutrino is emitted when the plasma temperature exceeds this threshold, we assume it interacts and discard it; otherwise, we retain it. We define this threshold as the \emph{energy-dependent decoupling temperature}.

Given that this effect provides only a small correction, we employ an approximate method sufficient to determine the correct order of magnitude. This approach is significantly more precise than simply discarding all neutrinos emitted before standard neutrino decoupling.

A standard way to determine neutrino decoupling is to compare the neutrino scattering rate, $\Gamma$, with the Hubble expansion rate, $H$. The scattering rate is given by
\begin{equation}
    \Gamma = n\, \langle \sigma v \rangle,
\end{equation}
where $n$ is the number density, $\sigma$ is the cross section, $v$ is the velocity, and $\langle ... \rangle$ denotes a thermal average. For relativistic particles, we approximate $v \approx 1$ and use the Fermi approximation for the cross section, $\sigma \approx \frac{G_F^2}{\pi} s$, where $G_F$ is Fermi’s constant and $s$ is the Mandelstam variable. This approximation holds up to $s \sim 100$ GeV.

\begin{figure}[t]
    \centering
    \includegraphics[width=0.75\linewidth]{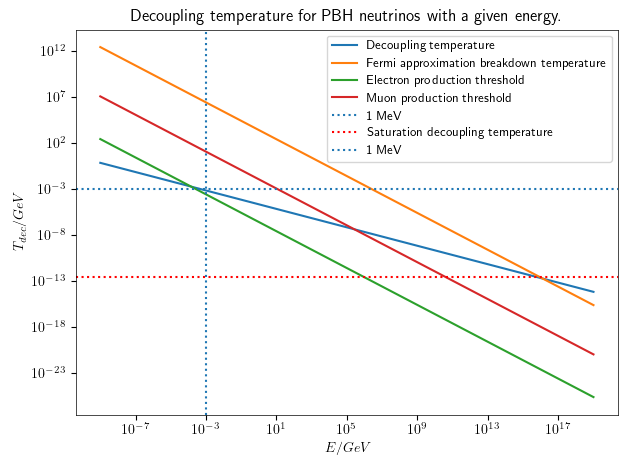}
    \caption{Energy-dependent decoupling temperatures where PBH neutrinos no longer interact with plasma neutrinos, assuming the energy of the plasma neutrino equals the plasma temperature (blue solid line). For comparison purposes, we also show the threshold for the interaction to produce an electron-positron pair (green solid line) or a muon-antimuon pair (red solid line), as well as the threshold where the Fermi approximation breaks down (orange solid line) and the saturation value reached after it breaks down (red dashed line). The blue dashed lines show an energy of 1 MeV, which is approximately the standard neutrino decoupling temperature.}
    \label{PBHdecouplingtemperatures}
\end{figure}

For a PBH neutrino of energy $E$ interacting with a plasma neutrino of energy $T$, assuming a head-on collision, we obtain $s = 4ET$. Substituting into $\Gamma$, we compare it with the Hubble rate, approximated as $H \approx T^2/M_P$, where $M_P = G^{-1/2}$ is the Planck mass. Setting $\Gamma \approx H$ gives
\begin{equation}
    T_{dec} = \sqrt{\frac{\pi}{4E M_P G_F^2}}.
\end{equation}
This shows that higher-energy neutrinos decouple at lower plasma temperatures. However, above $s \sim 100$ GeV, the cross section saturates at $\sigma \approx \frac{G_F^2}{\pi} M_W^2$, where $M_W$ is the W boson mass, leading to a constant decoupling temperature
\begin{equation}
    T_{dec} = \frac{\pi}{M_W^2 M_P G_F^2}.
\end{equation}
Figure~\ref{PBHdecouplingtemperatures} shows the energy-dependent decoupling temperature. Neutrinos emitted when the plasma temperature is above their corresponding $T_{dec}(E)$ are discarded.

Our results verify that neutrinos of $\sim 1$ MeV energy decouple at $\sim 1$ MeV, consistent with standard neutrino decoupling. For higher-energy neutrinos, decoupling occurs significantly later. Since PBHs typically emit neutrinos with energies above 1 MeV, they frequently interact before decoupling. Notably, many of these interactions have sufficient energy to produce electron-positron pairs, injecting energy into the photon plasma rather than the neutrino plasma. This supports our exclusion criterion, as such interactions redistribute energy across both sectors, preventing a significant contribution to $N_{\mathrm{eff}}$.

\section{Interactions of emitted unstable particles}\label{ap:MuonsPionsKaons}

 In this appendix, we want to briefly explain how we approximately calculate whether, for a given PBH mass $M$, emitted weakly-decaying unstable particles such as muons, charged pions or charged kaons \footnote{Neutral kaons also decay weakly, and do not really fit into the formalism of this appendix due to being neutral, but they only make up a small part of the spectrum, so we can neglect them.} decay before thermalizing or thermalize before decaying. To estimate the answer to this question, we focus on the case of muons, and we compare their decay rate with the scattering rate of muon-photon scattering, a process whose cross section can be deduced from that of Compton scattering \footnote{Due to lepton universality, it suffices to replace the electron mass by the muon mass.}, which is well-known \cite{Klein:1929lcc, Peskin:1995ev}. In the $s \gg m^2$ limit \footnote{This limit is not actually applicable in all of the regime we are considering. However, we have checked that using the full cross section (and using $s \sim 2ET + m^2$ instead of $s \sim 2ET$) essentially does not alter the result. Since using the full cross section makes the math more cumbersome and adds no new conceptual difficulties, we show the version in that limit here for the sake of simplicity.}, where $m$ is the muon mass, one has:

\begin{equation}
    \sigma = \frac{2\pi\alpha^2}{s} \ln\left( 1+\frac{s}{m^2} \right),
\end{equation}

where $m$ is the muon mass, $s$ is the square of the center-of-mass energy and $\alpha$ is the fine-structure constant. Knowing the cross section $\sigma$, we can calculate the scattering rate as $\Gamma_{\mu,\mathrm{scatter}} = n <\sigma v>$, where we take $v \sim 1$, and $n$ is the photon number density, which is related to the plasma temperature $T$ as:

\begin{equation}
    n = \frac{2 \zeta(3)}{\pi^2} T^3.
\end{equation}

To calculate the average cross section, we estimate that the center-of-mass energy is $s \sim 2ET$, where $E$ is the energy of the emitted particle, which we estimate as the Hawking temperature, $E \sim T_H$, and $T$ is the energy of the photon, which we approximate to be equal to the plasma temperature evaluated at the time of evaporation of PBHs of that mass. In reality, both the plasma temperature and the Hawking temperature vary throughout the PBH's lifetime, but we expect most of the spectrum to be emitted while the Hawking temperature is close to the initial value and the plasma temperature is close to its value at evaporation. This is an approximation, but it should yield the right order of magnitude, which is enough for the level of precision that we want here. The muon decay rate in its rest frame $\Gamma_{\mu,\mathrm{decay}}$ is well-known; we just need to divide it by a $\gamma$ factor due to time dilation to convert it to the lab frame (i. e. the CMB frame), with $\gamma = E/m$. Taking all of this into account, we get the following expression for the ratio between rates in that frame:

\begin{equation}
    \frac{\Gamma_{\mu, \mathrm{scatter}}}{\Gamma_{\mu,\mathrm{decay, lab}}} = T^2 \frac{2\zeta(3) \alpha^2}{\pi m \Gamma_{\mu,\mathrm{decay}}} \ln\left( 1+\frac{2ET}{m^2} \right).
    \label{eq:scattering_vs_decay_rates}
\end{equation}

By comparing that ratio with 1, we can know if muons thermalize or decay first. For pions and kaons, we assume that they behave like muons, but switching the mass and the decay rate to those corresponding to those particles. This is of course not correct, but it should yield the right order of magnitude, which is enough for the level of precision that we want here. With this procedure, we calculate the limits for the different regimes shown in subsection \ref{subsec:interactions}.

\section{Calibration of the spectra}\label{ap:SpectraCalibration}

A crucial step in our calculation is calibrating the spectra provided by \texttt{BlackHawk}. As noted in its manual, \texttt{BlackHawk} outputs spectra assuming a PBH number density of 1 PBH per comoving $\mathrm{cm}^3$. Comoving volume remains constant despite cosmic expansion, ensuring consistency in number densities across time.

We introduce the PBH abundance parameter $\alpha$, proportional to the PBH abundance, and defined such that $\alpha = 1$ corresponds to the \texttt{BlackHawk} spectra. While the scale factor evolution is inherently included in our integration, we require a global calibration factor to relate $\alpha$ to the commonly used parameter $\beta'$. This is achieved in two steps: first, relating $\alpha$ to the dark matter fraction parameter $f$, and then connecting $f$ to $\beta'$.

We normalize comoving volume to match today's physical volume, setting 1 comoving $\mathrm{cm}^3$ equal to 1 physical $\mathrm{cm}^3$ at present. Given that the current dark matter density is approximately $2 \times 10^{-27} \mathrm{kg}/\mathrm{m}^3$, which can be obtained from CLASS, we define $f=1$ as the case where PBHs constitute all of the dark matter. To simplify calibration, we ignore PBH evaporation and define $f$ as the fraction assuming PBHs do not evaporate—though not physically realistic, this definition is consistent for our purposes.

For a monochromatic PBH population of mass $M$, the PBH number density today is:
\begin{equation}
    n_{\rm PBH}(f=1) = \frac{\rho_{\rm PBH}(f=1)}{M}.
    \label{numberormassdensityPBHs}
\end{equation}
Since \texttt{BlackHawk} assumes $n_{\rm PBH}(\alpha = 1) = 1 \mathrm{cm}^{-3}$, we obtain a mass-dependent relation between $\alpha$ and $f$. The value of $f$ corresponding to a given $\alpha$ satisfies $n_{\rm PBH}(f(\alpha)) = \alpha \cdot 1 \mathrm{cm}^{-3}$, while also satisfying $n_{\rm PBH}(f(\alpha)) = f(\alpha) \cdot n_{\rm PBH}(f=1)$. Combining these with Eq.~\eqref{numberormassdensityPBHs}, we derive:
\begin{equation}
    f(\alpha) = \frac{\alpha M}{\rho_{\rm PBH}(f=1)} \cdot 1~\mathrm{cm}^{-3}.
\end{equation}
This relation allows us to determine $f$ for a given $\alpha$. To convert $f$ into $\beta'$, we use Eq.~(57) from Ref.~\cite{Carr:2020gox}
\begin{equation}
    f \approx 3.81 \times 10^8 \beta' \left(\frac{M}{M_\odot}\right)^{-1/2},
\end{equation}
where $M_\odot$ is the solar mass. Inverting this expression yields:
\begin{equation}
    \beta' \approx \left(\frac{M}{M_\odot}\right)^{1/2} \frac{f}{3.81\times 10^8}.
\end{equation}
Thus, this procedure provides a direct relation between $\alpha$ and $\beta'$. Since we obtain spectra from \texttt{BlackHawk} for $\alpha = 1$, we can derive results for any $\alpha$ by scaling them accordingly. By linking $\alpha$ to $\beta'$, we obtain results for arbitrary values of $\beta'$, enabling our program to place bounds on $\beta'$.

\end{appendices}


\bibliographystyle{JHEP}
\bibliography{Bibliography.bib}

\end{document}